# Optimal Reconstruction of the Hellings and Downs Correlation

Bruce Allen[*]

*Max Planck Institute for Gravitational Physics (Albert Einstein Institute),*
*Leibniz Universität Hannover, Callinstrasse 38, D-30167, Hannover, Germany*

Joseph D. Romano[†]

*Department of Physics and Astronomy, University of Texas Rio Grande Valley,*
*One West University Boulevard, Brownsville, Texas 78520, USA*

Pulsar timing arrays (PTAs) detect gravitational waves (GWs) via the correlations they create in the arrival times of pulses from different pulsars. The mean correlation, a function of the angle between the directions to two pulsars, was predicted in 1983 by Hellings and Downs (HD). Observation of this angular pattern is crucial evidence that GWs are present, so PTAs "reconstruct the HD curve" by estimating the correlation using pulsar pairs separated by similar angles. Several studies have examined the amount by which this curve is expected to differ from the HD mean. The variance arises because (a) a finite set of pulsars at specific sky locations is used, (b) the GW sources interfere, and (c) the data are contaminated by noise. Here, for a Gaussian ensemble of sources, we predict that variance by constructing an optimal estimator of the HD correlation, taking into account the pulsar sky locations and the frequency distribution of the GWs and the pulsar noise. The variance is a ratio: the numerator depends upon the pulsar sky locations, and the denominator is the (effective) number of frequency bins for which the GW signal dominates the noise. In effect, after suitable combination, each such frequency bin gives an independent estimate of the HD correlation.

*Introduction*—As pulsar timing arrays (PTAs) work towards $5\sigma$ detections of gravitational waves (GWs) [1–4], there is growing interest in different aspects of the underlying physics. This includes potential GW sources, mechanisms that influence pulsar rotation, and the propagation, detection, and analysis of electromagnetic pulses. These, in turn, inform the data analysis.

If GWs had large amplitudes, their effects on the arrival times of pulses from a single pulsar would be directly visible. Early work [5] set upper limits on the GW amplitude using individual pulsars, but it is now known that GW effects are small compared to pulsar timing noise. So, to detect GWs, PTAs search for GW-induced correlations in the arrival times of pulses from *different* pulsars.

The correlation $-1 \leq \frac{3}{2}\mu \leq 1$ in pulsar timing residuals is a function of the angle $\gamma \in [0,\pi]$ between the directions to pulsars. These "spatial" or "angular" correlations can be expressed as a sum of Legendre polynomials

$$\mu(\gamma) = \sum_l c_l P_l(\cos\gamma). \qquad (1)$$

The coefficients $c_l$ may be estimated from the data: 100 pulsars would give values of $\mu$ at 4950 angles $\gamma > 0$.

The expected pattern of correlation $\mu_u(\gamma) = \langle\mu(\gamma)\rangle$ for a stationary, isotropic, and unpolarized gravitational wave background (GWB) was predicted in 1983 by Hellings and Downs (HD). This "HD curve" [6] is plotted in the top panel of Fig. 1 and is

$$\mu_u(\gamma) = \frac{1}{3} + \frac{1-\cos\gamma}{2}\left[\ln\left(\frac{1-\cos\gamma}{2}\right) - \frac{1}{6}\right]. \qquad (2)$$

The expected coefficients $\langle c_0 \rangle = 0$, $\langle c_1 \rangle = 0$, and

$$\langle c_l \rangle = (2l+1)/((l+2)(l+1)l(l-1)) \quad \text{for } l \geq 2 \qquad (3)$$

are computed in [7–9]. [The correlation is doubled to $3\mu_u(0) = 1$ for pulsars that are closer together than the typical GW wavelength ([10], Appendix C.2).] Detection of the HD curve is crucial evidence that the pulsar arrival time fluctuations are due to GWs [11].

How closely will the correlations in our (realization of the) Universe follow this curve? Even if the noise is small, deviations occur because of (i) pulsar variance and (ii) cosmic variance. The first arises because observations are carried out with a finite set of pulsars at specific sky locations [12]. The second arises because our Universe has discrete GW sources with specific frequencies, sky locations, and other parameters. Interference between these sources means that even if pulsar variance is eliminated by using many pulsars [10,13,14], the pulsar-averaged correlation curve [15] will still differ from the HD curve.

[*]Contact author: bruce.allen@aei.mpg.de
[†]Contact author: joseph.romano@utrgv.edu

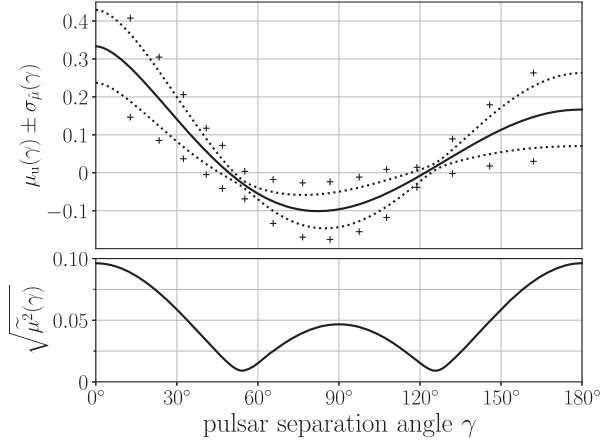

FIG. 1. Top: the black line is the HD curve $\mu_u(\gamma)$ of (2). The "+" symbols are the predicted $\pm\sigma_{\hat{\mu}} = \pm 1\sigma_G$ deviations of (30), for a reconstruction using NANOGrav's [2] 15 angular bins and 67 pulsars, if $N_{\text{freq}} = 1$. The dotted line is the same prediction in the limit of an infinite number of pulsars (7). Bottom: square root of the function $\widetilde{\mu^2}(\gamma)$ of (5).

We quantify such deviations via the variance

$$\sigma_\mu^2(\gamma) \equiv \langle \mu(\gamma)^2 \rangle - \langle \mu(\gamma) \rangle^2, \qquad (4)$$

noting that its value and meaning depend upon the statistical ensemble used for the angle brackets.

*Previous work*—The question was first addressed by Roebber and Holder [8]. They assumed that the GWB is described by an isotropic and unpolarized Gaussian ensemble of sources radiating GWs at a single frequency, and that noise-free measurements are carried out using an infinite number of pulsars. While not given in this form (see [10], Appendix C6, and [16]), they obtain

$$\sigma_\mu^2(\gamma) = \widetilde{\mu^2}(\gamma) \equiv \sum_l \frac{\langle c_l \rangle^2}{2l+1} P_l^2(\cos\gamma), \qquad (5)$$

whose square root is shown in the bottom panel of Fig. 1. This follows from a sky-map decomposition of the pulsar redshifts into spherical harmonics $Y_{lm}(\Omega_p)$, where $\Omega_p$ is the pulsar location on the two-sphere [9]. For fixed $l$, the $c_l$ are $\chi^2$ distributed with $(2l+1) \times 2$ degrees of freedom; the ratio of the variance to the squared mean is $1/(2l+1)$ as seen in (5). This variance was also found in [10] [with a closed form for $\widetilde{\mu^2}(\gamma)$, Eq. (G11)], where it was shown to arise from interference between GW sources.

More recent work examines the effects of pulsar and cosmic variance. Most [10,14,16,17] use the "pulsar averaging" technique of [15] to eliminate pulsar variance in the many-pulsar limit. The resulting cosmic variance has been computed for three GWB models: a Gaussian ensemble with an arbitrary spectrum [10,16] and discrete-source ensembles containing $N$ circularly or elliptically polarized GW sources [10,14,17]. Cosmic variance only arises if the GW sources interfere; the Roebber and Holder result is recovered in the limit of a high density of weak monochromatic GW sources. Other work [12] shows how to optimally combine the correlations of a finite set of pulsars at specific sky locations to estimate the HD correlation. The variance of this estimator reduces to the previously studied cosmic variance in the limit of large numbers of uniformly distributed pulsars, see e.g., [12], Fig. 7.

*Summary*—In this paper, we construct the best possible estimator $\hat{\mu}$ of the HD correlation for a Gaussian ensemble using a finite set of pulsar pairs in an angular bin at angle $\gamma$. The estimator (23) weights the data by frequency, thus including nonzero-lag information. Our main result: the variance of the estimator is

$$\sigma_{\hat{\mu}}^2 = \frac{\sigma_G^2}{N_{\text{freq}}}. \qquad (6)$$

Here, $\sigma_G^2$ (31) is a geometric quantity, determined entirely by the pulsar locations, and $N_{\text{freq}} \geq 0$ given by (32) is the (effective, noninteger) number of frequency bins for which the GWB dominates the pulsar spin/timing noise. Both $\sigma_G^2$ and $N_{\text{freq}}$ depend upon $\gamma$. Figure 2 illustrates how $\sigma_{\hat{\mu}}^2$ and $N_{\text{freq}}$ vary as the level of pulsar noise changes, for a simple numerical example. For a narrow angular bin containing many pulsar pairs uniformly distributed on the sky, $\sigma_G^2 \to \widetilde{\mu^2}(\gamma)$ and (6) becomes

$$\sigma_{\hat{\mu}}^2 = \frac{1}{N_{\text{freq}}} \widetilde{\mu^2}(\gamma) \Leftrightarrow \sigma_{\hat{c}_l}^2 = \frac{1}{N_{\text{freq}}} \frac{\langle c_l \rangle^2}{2l+1}, \qquad (7)$$

extending Roebber and Holder's $N_{\text{freq}} = 1$ result (5). Note that (7) for $\sigma_{\hat{c}_l}^2$ holds only if $N_{\text{freq}}$ is independent of $\gamma$ [18].

*Derivation*—In the Earth-pulsar neighborhood, far from any GW sources, GWs are described by a plane-wave expansion [10], Eq. (C1). The transverse traceless synchronous metric perturbations arising from GWs are

$$h_{\mu\nu}(t,\boldsymbol{x}) = \sum_A \int df \int d\Omega\, h_A(f,\Omega) e_{\mu\nu}^A(\Omega) e^{2\pi i f(t - \boldsymbol{\Omega}\cdot\boldsymbol{x})}, \qquad (8)$$

where the spatial coordinate $\boldsymbol{x} = 0$ at Earth and time $t$ is measured there. In (8), the GW frequency $f \in \mathfrak{R}$, and the unit vector $\boldsymbol{\Omega}$ is the GW propagation direction, touching the unit two-sphere at spherical polar coordinates $\Omega = (\theta, \phi)$.

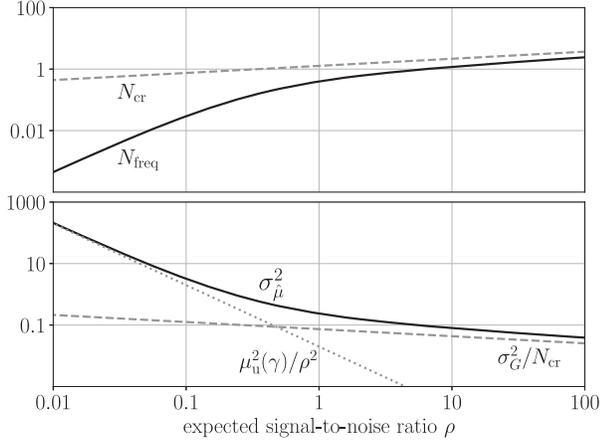

FIG. 2. A simple example with fixed GWB power spectrum $H(f) \propto f^{-7/3}$, where the level of pulsar noise $P(f) \propto f^2$ (same for all pulsars) is varied. The exponent $-7/3$ corresponds to binary inspiral and the exponent 2 corresponds to a white spectrum of timing-residual noise. The horizontal axis is the expected SNR $\rho$ for an angular bin at $\gamma = 30°$ containing three pulsar pairs. Top: $N_{cr}$ is the index of the frequency bin where the GWB and pulsar noise power spectra cross. $N_{freq}$, given by (32), is the effective number of frequency bins for which the GWB dominates the noise. Bottom: variance $\sigma_{\hat{\mu}}^2 = \sigma_G^2/N_{freq}$ of the reconstructed HD curve, where $\sigma_G^2$ is given by (31). For small SNR, $\sigma_{\hat{\mu}}^2 \approx \mu_u^2(\gamma)/\rho^2$ is noise dominated. For large SNR, $\sigma_{\hat{\mu}}^2 \approx \sigma_G^2/N_{cr}$ is cosmic variance dominated.

The infinitesimal area on the sphere is $d\Omega = \sin\theta d\theta d\phi$, the spatial coordinate indices $\mu, \nu \in \{x, y, z\}$, the polarization label $A \in \{+, \times\}$, the polarization tensors $e_{\mu\nu}^+$ and $e_{\mu\nu}^\times$ depend upon the GW direction, and $h_+$ and $h_\times$ are arbitrary complex functions satisfying $h_A^*(f, \Omega) = h_A(-f, \Omega)$, since $h_{\mu\nu}$ is real.

Consider a pulsar $a$ at distance $L_a > 0$ from Earth in direction $\Omega_a$, so $x_a = L_a \Omega_a$. The redshift of the pulsar's frequency at time $t$ on Earth

$$Z_a(t) = z_a(t) + n_a(t) \qquad (9)$$

is the sum of a GW-induced term $z_a(t)$ and a pulsar noise term $n_a(t)$, which can be written in terms of Fourier transforms

$$z_a(t) = \sum_A \int df \int d\Omega\, h_A(f, \Omega) F_a^A(\Omega) \tau(f, L_a\Omega_a, \Omega) e^{2\pi i f t},$$

$$n_a(t) = \int df\, \tilde{n}_a(f) e^{2\pi i f t}. \qquad (10)$$

The pulsar "antenna pattern" for polarization $A$ is

$$F_a^A(\Omega) = \frac{1}{2}\frac{\Omega_a^\mu \Omega_a^\nu e_{\mu\nu}^A(\Omega)}{1 + \Omega \cdot \Omega_a}, \qquad (11)$$

with the Einstein summation convention applying to $\mu$ and $\nu$. The factor $\tau$ forms the difference between Earth and pulsar terms $\tau(f, L_a\Omega_a, \Omega) = 1 - e^{-2\pi i f L_a(1 + \Omega \cdot \Omega_a)}$; see [13].

This analysis could also start with pulsar timing residuals, which are the time integrals of the redshifts. This does not change the value of the estimator of the HD correlation [19], or its ensemble mean and variance.

Expressions (8), (9), and (10) describe the GWs and the redshift over intervals (millions of years) much shorter than the Hubble time; PTAs observe a snapshot of the redshift (9) over a time interval $T$ of order decades. Assume that the redshift is observed over a time period $t \in [-T/2, T/2]$. We express this as a Fourier sum

$$Z_a(t) = \sum_j Z_a^j e^{2\pi i f_j t} \quad \text{for } t \in [-T/2, T/2], \qquad (12)$$

where the frequencies $f_j \equiv j/T$ are integer multiples of $1/T$, and the sum has $j \in -N_{bin}, \ldots, -1, 1, \ldots, N_{bin}$. The number of frequency bins $N_{bin}$ must be large enough that (i) the Nyquist frequency $N_{bin}/T$ lies well above the largest GWB-dominated frequency and (ii) $T/2N_{bin}$ is much less than the time between successive pulsar-timing observations. For quantities that carry both pulsar and frequency indices, we put pulsar indices $a, b, c, d, e, f$ down and frequency indices $j, k, \ell, m, r, s$ up.

In what follows, we assume that a timing model has been fit to the data [20]. This removes certain functions of time (e.g., linear trends in timing residuals, or constants in redshift) from the data set, but is gracefully accommodated by our analysis. The only effect is to modify the power spectra $H(f)$ and $P_a(f)$ defined by (14), thus modifying the correlations between the different frequency bins in the corresponding matrices $H_{jk}$ and $P_a^{jk}$ defined by (16); see [21].

Multiplying the expressions in (10) by $T^{-1}e^{-2\pi i f_k t}$ and integrating over $t \in [-T/2, T/2]$, and then doing the same to (12) gives the redshift amplitude in the $k$th frequency bin, $Z_a^k \equiv z_a^k + n_a^k$, with

$$z_a^k = \sum_A \int df \int d\Omega\, h_A(f, \Omega) F_a^A(\Omega) \tau(f, L_a\Omega_a, \Omega)$$
$$\times \text{sinc}(\pi(f - f_k)T),$$

$$n_a^k = \int df\, \tilde{n}_a(f) \text{sinc}(\pi(f - f_k)T). \qquad (13)$$

Since (10) are real, $z_a^{k*} = z_a^{-k}$, $n_a^{k*} = n_a^{-k}$, and $Z_a^{k*} = Z_a^{-k}$. The $\text{sinc}(x) \equiv (\sin x)/x$ creates sidelobes from sources whose frequencies are not integer multiples of $1/T$. Hence, because $T\text{sinc}(\pi(f - f_k)T) \neq \delta(f - f_k)$, the Fourier coefficients $n_a^k$ and $z_a^k$ are *not* given by the (continuous) Fourier transforms evaluated at $f = f_k$.

Since the pulsar noises $n_a(t)$ are unknown, and the parameters (sky positions, frequencies, amplitudes, etc.) of the GW sources contributing to $h_A(f, \Omega)$ are unknown, we cannot predict the redshifts. Instead, we give a statistical description. Assuming that the GWs and pulsar noise arise independently from incoherent sums of many weak processes, the central limit theorem applies, so both may be described by stationary Gaussian ensembles.

Let $h_A(f, \Omega)$ and $\tilde{n}_a(f)$ be representative functions drawn from stationary Gaussian ensembles describing an isotropic unpolarized GWB and uncorrelated pulsar noise. Then, using angle brackets to denote averages over this ensemble, the Gaussian process is fully defined by its first $\langle h_A(f, \Omega)\rangle = \langle \tilde{n}_a(f)\rangle = 0$ and second moments

$$\langle h_A(f, \Omega) h_{A'}^*(f', \Omega')\rangle = \delta_{AA'}\delta(f-f')\delta^2(\Omega, \Omega')H(f),$$
$$\langle \tilde{n}_a(f))\tilde{n}_b^*(f')\rangle = \delta_{ab}\delta(f-f')P_a(f),$$
$$\langle h_A(f, \Omega))\tilde{n}_a^*(f')\rangle = 0, \qquad (14)$$

where $H(f) = H(-f) \geq 0$ and $P_a(f) = P_a(-f) \geq 0$ are real power spectra of the GWB and noise ([10], Eqs. (C5–7), relates $H(f)$ to the GW energy density and other observables). The $\delta_{ab}$ term in (14) reflects our assumption that the noise in the different pulsars is statistically independent. All higher moments can be computed from the first and second moments via Isserlis's theorem [22].

The Fourier coefficients $Z_a^k$ are linear combinations of $h_A(f, \Omega)$ and $\tilde{n}_a(f)$, so they are also Gaussian random variables. Their first moments vanish $\langle Z_a^k\rangle = 0$, and from (13) and (14) their second moments are

$$\langle Z_a^j Z_b^{k*}\rangle = H_{jk}\mu_{ab} + P_a^{jk}\delta_{ab}. \qquad (15)$$

Here, $\boldsymbol{H} \equiv H_{jk}$ and $\boldsymbol{P}_a \equiv P_a^{jk}$ are real bisymmetric $2N_{\text{bin}} \times 2N_{\text{bin}}$ matrices with rows and columns indexed by frequency bin. They are defined by the relations

$$H_{jk} \equiv 4\pi \int df\, H(f)\text{sinc}(\pi(f-f_j)T)\text{sinc}(\pi(f-f_k)T),$$
$$P_a^{jk} \equiv 4\pi \int df\, P_a(f)\text{sinc}(\pi(f-f_j)T)\text{sinc}(\pi(f-f_k)T).$$
$$(16)$$

$\boldsymbol{H}$ and $\boldsymbol{P}_a$ have non-negative eigenvalues, and their matrix inverses are denoted for example by $\boldsymbol{H}^{-1}$; if $\det(\boldsymbol{H}) = 0$, then $\boldsymbol{H}^{-1}$ denotes the (Moore-Penrose) pseudoinverse. Since $H_{jk} = H_{kj} = H_{-k,-j}$, both $\boldsymbol{H}$ and $\boldsymbol{H}^{-1}$ are reflection invariant across *either* diagonal; the same holds for $\boldsymbol{P}_a$ and $\boldsymbol{P}_a^{-1}$. (For the Fig. 2 example, the integrals are restricted to $f \in [1/2T, 30/T]$, because fitting to a timing model removes contributions from frequencies below $f \sim 1/T$ [20,23].)

The object $\boldsymbol{\mu} \equiv \mu_{ab}$ that appears in (15) is

$$\boldsymbol{\mu} \equiv \mu_{ab} \equiv \mu_u(\gamma_{ab})(1 + \delta_{ab}). \qquad (17)$$

It has indices labeled by pulsars $a$ and $b$, and its entries are the values of the HD curve at angle $\gamma_{ab}$, doubled if $a$ and $b$ are the same. The angle $\gamma_{ab} \in [0, \pi]$ between the lines of sight to $a$ and $b$ is defined by $\cos\gamma_{ab} = \boldsymbol{\Omega}_a \cdot \boldsymbol{\Omega}_b$.

To obtain (15) and (17), we used the definition of the Hellings and Downs curve

$$\mu_u(\gamma_{ab}) \equiv \frac{1}{4\pi}\sum_A \int d\Omega\, F_a^A(\Omega)F_b^A(\Omega), \qquad (18)$$

and the reasoning given in [10], Appendix C2, to replace $\tau(f, L_a\boldsymbol{\Omega}_a, \Omega)\tau^*(f, L_b\boldsymbol{\Omega}_b, \Omega) \to 1 + \delta_{ab}$ within integrals over frequency $f$ and direction $\Omega$.

Later, we need the real covariance matrix

$$\begin{aligned}\mathcal{C}_{ab,cd}^{jk,\ell m} &\equiv \langle Z_a^j Z_b^k Z_c^{\ell*} Z_d^{m*}\rangle - \langle Z_a^j Z_b^k\rangle\langle Z_c^{\ell*} Z_d^{m*}\rangle\\
&= \langle Z_a^j Z_c^{\ell*}\rangle\langle Z_b^k Z_d^{m*}\rangle + \langle Z_a^j Z_d^{m*}\rangle\langle Z_b^k Z_c^{\ell*}\rangle\\
&= \mu_{ac}\mu_{bd}H_{j\ell}H_{km} + \mu_{ad}\mu_{bc}H_{jm}H_{k\ell} + \delta_{ac}\mu_{bd}P_a^{j\ell}H_{km}\\
&\quad + \delta_{ad}\mu_{bc}P_a^{jm}H_{k\ell} + \mu_{ac}\delta_{bd}H_{j\ell}P_b^{km} + \mu_{ad}\delta_{bc}H_{jm}P_b^{k\ell}\\
&\quad + \delta_{ac}\delta_{bd}P_a^{j\ell}P_b^{km} + \delta_{ad}\delta_{bc}P_a^{jm}P_b^{k\ell},\end{aligned} \qquad (19)$$

where the second equality follows from Isserlis's theorem [22], and the third from (15). For reasons described below, we require the part of $\mathcal{C}_{ab,cd}^{jk,\ell m}$ which is symmetric in both $jk$ and $\ell m$:

$$\boldsymbol{C} \equiv C_{ab,cd}^{jk,\ell m} \equiv \mathcal{C}_{ab,cd}^{(jk),(\ell m)}, \qquad (20)$$

where the round brackets denote symmetrization, e.g., $Q_{(jk)} \equiv (Q_{jk} + Q_{kj})/2$. It is easy to see that the noise-free component of $C_{ab,cd}^{jk,\ell m}$ factors into

$$C_{ab,cd}^{jk,\ell m}\bigg|_{\boldsymbol{P}=0} = G_{ab,cd}H_{j(\ell}H_{m)k}. \qquad (21)$$

The "geometry" factor depends on the pulsar directions,

$$\boldsymbol{G} \equiv G_{ab,cd} \equiv \mu_{ac}\mu_{bd} + \mu_{ad}\mu_{bc}, \qquad (22)$$

and plays an important role in [12].

To estimate the HD correlation at angle $\gamma$, we use $N_{\text{pair}}$ pulsar pairs $ab$ lying in an angular bin around $\gamma$. Following [12], we use the notation $ab \in \gamma$ to denote this set of pulsar pairs; autocorrelations are excluded, so $a < b$. See [12], Fig. 2, for a helpful illustration with $N_{\text{pair}} = 3$.

The estimator $\hat{\mu}$ is a general linear combination

$$\hat{\mu} \equiv \sum_{ab \in \gamma} \sum_{j,k} W_{ab}^{jk} Z_a^j Z_b^k \qquad (23)$$

of redshift cross-products. The weights $W \equiv W_{ab}^{jk}$ are set by requiring that $\hat{\mu}$ (i) is unbiased, (ii) minimizes the variance among universes drawn from the Gaussian ensemble, and (iii) is real, so $W_{ab}^{jk*} = W_{ab}^{-j,-k}$.

Equation (23) is more general than the estimator of [12], which is a linear combination of the zero-lag correlations $\rho_{ab} \equiv (1/T) \int_{-T/2}^{T/2} dt\, Z_a(t) Z_b(t) = \sum_j Z_a^j Z_b^{j*}$. Those estimators have weights $W_{ab}^{jk}$, which vanish off the antidiagonal $j = -k$ and are $j$ (frequency) independent. The more general form (23) allows us to further reduce the variance, thus improving the estimator.

The optimal weights $W_{ab}^{jk}$ are found as in [12], Sec. 3A. From (15), the ensemble average of the estimator (23) is

$$\langle \hat{\mu} \rangle = \sum_{ab \in \gamma} \sum_{j,k} W_{ab}^{jk} \mu_{ab} \overline{H}_{jk} = \mu_u(\gamma), \qquad (24)$$

where the noise terms $P$ are absent because $a < b$. The final equality above is a constraint: normalizing to $\mu_u(\gamma)$ ensures that $\hat{\mu}$ is unbiased. We have defined $\bar{H} \equiv \bar{H}_{jk} \equiv H_{j,-k}$; it is symmetric in $jk$ since $H_{j,-k} = H_{-j,k} = H_{k,-j}$. Since (24) does not constrain the part of $W_{ab}^{jk}$ antisymmetric in $jk$, we set that to zero: $W_{ab}^{jk} = W_{ab}^{kj} = W_{ab}^{(jk)}$.

The variance (4) of $\hat{\mu}$ is $\sigma_{\hat{\mu}}^2 \equiv \langle |\hat{\mu}|^2 \rangle - |\langle \hat{\mu} \rangle|^2$, so from (19), (20), and (23),

$$\sigma_{\hat{\mu}}^2 = \sum_{ab \in \gamma} \sum_{cd \in \gamma} \sum_{j,k} \sum_{\ell,m} W_{ab}^{jk} C_{ab,cd}^{jk,\ell m} W_{cd}^{\ell m*}. \qquad (25)$$

The right-hand side of (25) motivates the introduction of an inner product between weights $A \equiv A_{ab}^{jk}$ and $B \equiv B_{cd}^{\ell m}$:

$$(A,B) \equiv A^t C B^* \equiv \sum_{ab \in \gamma} \sum_{cd \in \gamma} \sum_{j,k} \sum_{\ell,m} A_{ab}^{jk} C_{ab,cd}^{jk,\ell m} B_{cd}^{\ell m*} \qquad (26)$$

This inner product is positive definite if the pulsar positions are generic, $H$ and $P_a$ have nonzero eigenvalues, and $A$ and $B$ are symmetric in the frequency indices.

The covariance $C$ is a rank $N_{\text{bin}}(2N_{\text{bin}} + 1)N_{\text{pair}}$ square symmetric matrix with $4N_{\text{bin}}^2 N_{\text{pair}}$ rows and columns. It has $N_{\text{bin}}(2N_{\text{bin}} - 1)N_{\text{pair}}$ eigenvectors antisymmetric in $jk$ with eigenvalue zero. Its rank $N_{\text{bin}}(2N_{\text{bin}} + 1)N_{\text{pair}}$ pseudoinverse $C^{-1} \equiv (C^{-1})_{ab,cd}^{jk,\ell m}$ satisfies

$$\sum_{cd \in \gamma} \sum_{\ell,m} (C^{-1})_{ab,cd}^{jk,\ell m} C_{cd,ef}^{\ell m, rs} = 2\delta_{e(a}\delta_{b)f}\delta_{r(j}\delta_{k)s}$$

$$= \delta_{ae}\delta_{bf}\delta_{r(j}\delta_{k)s}, \qquad (27)$$

where the last equality holds if $a < b$ and $e < f$.

It is helpful to introduce weights $V \equiv C^{-1}\mu\bar{H}$

$$V \equiv V_{ab}^{jk} \equiv \sum_{cd \in \gamma} \sum_{\ell,m} (C^{-1})_{ab,cd}^{jk,\ell m} \mu_{cd} \bar{H}_{\ell m}. \qquad (28)$$

These are symmetric in $jk$ and real, satisfying $V_{ab}^{jk*} = V_{ab}^{jk} = V_{ab}^{-j,-k}$.

To find the weights $W$ defining the minimum-variance estimator $\hat{\mu}$, use the inner product (26) to write

$$\langle \hat{\mu} \rangle = (W, V) \quad \text{and} \quad \sigma_{\hat{\mu}}^2 = (W, W). \qquad (29)$$

The first equality follows from (24), (26), and (28), and the second from (25) and (26). Minimizing the variance subject to the normalization constraint $\langle \hat{\mu} \rangle = \mu_u(\gamma)$ is equivalent to minimizing the ratio $(W, W)/(W, V)^2$ by maximizing the denominator. Since (26) is positive definite, this implies that $W$ is proportional to $V$. The correctly normalized solution is $W = \mu_u(\gamma) V/(V, V)$, which is similar in form to optimal signal-detection statistics [24–26]. It differs from existing optimal statistic estimators [27] because those only minimize the variance in the weak-signal limit, whereas ours does so for any ratio of signal to noise.

The variance of the estimator follows from (26), (27), (28), and (29), and is

$$\sigma_{\hat{\mu}}^2 = (W, W) = \frac{\mu_u^2(\gamma)}{(V, V)} = \frac{\sigma_G^2}{N_{\text{freq}}}. \qquad (30)$$

Here, we have defined

$$\sigma_G^2 \equiv \frac{\mu_u^2(\gamma)}{2\mu^t G^{-1} \mu}, \qquad (31)$$

$$N_{\text{freq}} \equiv \frac{(\mu \bar{H})^t C^{-1}(\mu \bar{H})}{2\mu^t G^{-1} \mu}. \qquad (32)$$

The $N_{\text{pair}} \times N_{\text{pair}}$ square symmetric matrix $G^{-1}$ satisfies

$$\sum_{cd \in \gamma} G_{ab,cd}^{-1} G_{cd,ef} \equiv \delta_{ae}\delta_{bf} + \delta_{af}\delta_{be} = \delta_{ae}\delta_{bf}, \qquad (33)$$

where the last equality holds for $a < b$ and $e < f$.

In (31) and (32), $\mu$ and $\mu\bar{H}$ are column vectors with respective dimensions $N_{\text{pair}}$ and $4N_{\text{bin}}^2 N_{\text{pair}}$. The quantity $\sigma_G^2$ (shown in Fig. 1 for NANOGrav's [2] 67 pulsars and 15 angular bins) is the one-frequency-bin cosmic variance found in [12] for a specific finite set of pulsar pairs. Below, we show that $N_{\text{freq}}$ is the effective number of frequency bins for which the GW signal dominates the noise; see Fig. 2.

Equation (30) is our main result. It is similar to the variance found in [12], but smaller because our new estimator (23) combines information from different frequencies. Furthermore, as an optimal estimator, it is independent of the form of the data (e.g., timing residuals versus redshifts), because the powers of $f$ relating them cancel out of the products of $H$, $P_a$, $H^{-1}$, and $P_a^{-1}$, see [19].

If the angular bin is narrow, then $\mu \approx \mu_u(\gamma)\mathbb{1}$, where $\mathbb{1} = (1, ..., 1)^t$ is a column vector containing $N_{\text{pair}}$ ones. For this narrow angular bin, discrete pulsar pair case

$$\sigma_G^2 = \frac{1}{2\mathbb{1}^t G^{-1} \mathbb{1}}. \quad (34)$$

If there are many pulsar pairs in a bin at angle $\gamma$, uniformly distributed on the sky, then [12] shows that $(\mathbb{1}^t G^{-1} \mathbb{1})^{-1} \to 2\widetilde{\mu^2}(\gamma)$. For this case, $\sigma_G^2 \to \widetilde{\mu^2}(\gamma)$, from which our other key result (7) follows from (30).

To see that $N_{\text{freq}}$ is the effective number of frequency bins for which the GWB dominates pulsar noise, consider the following simple model: (i) The pulsar noise spectrum is pulsar independent, so $P_a^{jk} = P_{jk}$. (ii) Below some frequency bin $N_{\text{cr}}$, there is only a GWB, so $P_{jk} = 0$ if $|j|$ or $|k| \leq N_{\text{cr}}$. (iii) Above frequency $N_{\text{cr}}/T$, there is only pulsar noise, so $H_{jk} = 0$ if $|j|$ or $|k| > N_{\text{cr}}$. This is a simplified version of the more realistic case where a "red" GWB power spectrum crosses "white" pulsar-noise power spectra at frequency bin $N_{\text{cr}}$.

With these assumptions, the matrix $H_{jk} + P_{jk}$ is block diagonal. Since the "signal part" and the "noise part" act on orthogonal subspaces, $C^{-1}$ is given by

$$(C^{-1})_{ab,cd}^{jk,\ell m} = G_{ab,cd}^{-1} H_{\ell(j}^{-1} H_{k)m}^{-1} + 2\delta_{c(a}\delta_{b)d} P_{\ell(j}^{-1} P_{k)m}^{-1}, \quad (35)$$

where $H^{-1}$ and $P^{-1}$ denote pseudoinverses, with respective ranks $2N_{\text{cr}}$ and $2N_{\text{bin}} - 2N_{\text{cr}}$. Equation (35) can be verified by inserting it and the symmetric part of (19) into (27), since conditions (ii) and (iii) above imply that any contraction of $H$ with $P$ vanishes. From (35),

$$(\mu\bar{H})^t C^{-1} (\mu\bar{H}) = \mu^t G^{-1} \mu \, \text{Tr}(\bar{H} H^{-1} \bar{H} H^{-1}), \quad (36)$$

so (32) implies that

$$N_{\text{freq}} = \frac{1}{2} \sum_{j,k} \sum_{\ell,m} \bar{H}_{jk} \bar{H}_{\ell m} H_{j\ell}^{-1} H_{km}^{-1} = N_{\text{cr}}, \quad (37)$$

which is half the rank of $H$ or of $H^{-1}$. The final equality in (37) follows by writing $\bar{H}_{jk} = H_{j,-k}$, relabeling $k \to -k$ and $m \to -m$, and using $H_{-k,-m}^{-1} = H_{km}^{-1}$.

If the GWB is weak compared to the noise, then $N_{\text{freq}} \to 0$. Let $S/\sqrt{\langle S^2 \rangle_0}$ denote the signal-to-noise ratio (SNR) for the pulsar pairs in the angular bin. Here, $S$ is the optimal cross-correlation GWB *detection statistic*, defined by the right-hand side of (23) with (different) weights $w$, and $\langle S^2 \rangle_0$ is its variance in the *absence* of a GWB. The weights are set to maximize the square of the expected SNR in the presence of a GWB, $\rho^2 \equiv \langle S \rangle^2 / \langle S^2 \rangle_0$. Via the same reasoning leading to (30), one obtains

$$\rho^2 = \frac{(w, C_0^{-1} \bar{H}\mu)_0^2}{(w, w)_0} = (\mu\bar{H})^t C_0^{-1} (\mu\bar{H}), \quad (38)$$

where the optimal weights are $w = C_0^{-1} \bar{H}\mu$. The subscripts 0 indicate that $H = 0$ in the covariance matrix (19) and for $C$ inside the inner product (26).

If the pulsar noise $P$ is large then $C \to C_0$ and $\rho^2$ is small. It follows from (32) that $N_{\text{freq}} \approx \rho^2 \sigma_G^2 / \mu_u^2(\gamma)$ is small and that $\sigma_\mu^2 = \sigma_G^2 / N_{\text{freq}} \approx \mu_u^2(\gamma) / \rho^2$ is large. For this case, the variance of the optimal estimator of the HD correlation is large, and dominated by pulsar noise rather than by cosmic variance; see Fig. 2. [Note: for pulsar-independent noise, $C_0^{-1}$ is given by the second term on the right-hand side of (35), so (38) gives a squared expected SNR $\rho^2 = \mu_u^2(\gamma) N_{\text{pair}} \text{Tr}(HP^{-1} HP^{-1})$.]

Current HD curve reconstructions are noise dominated. For example, for NANOGrav's 15 angular bins and 67 pulsars, the uncertainties (error bars) of Fig. 1(c) of [2] correspond to values of $N_{\text{freq}}$ in the range from 0.73 to 0.29, with a mean of 0.47 and a standard deviation of 0.13. Our optimal estimator should have comparable or larger values of $N_{\text{freq}}$.

*Conclusion*—Roebber and Holder [8], end of Sec. 4, write that "separate frequency bins can be considered as independent realizations of the same map." The word "considered" is needed: since $H$ is nondiagonal, the maps are correlated and not independent. This is generic to PTAs, whose observational time spans are much shorter than the coherence time of their GW sources.

Our calculation proves that this intuition is correct. It gives a rigorous definition of the number of signal-dominated frequency bins $N_{\text{freq}}$, showing that each provides an independent estimator of the HD correlation. Optimally combining the data reduces the total variance in proportion to $N_{\text{freq}}$. It is satisfying that these results also hold for finite numbers of pulsars at specific sky locations, not just in the infinite-pulsar limit, and that the HD estimator constructed from redshifts and the HD estimator constructed from timing residuals have identical values and variances [19].

The variance of the harmonic coefficients $c_l$ in (1) has similar form and behavior; the dependence of $N_{\text{freq}}$ on $l$ for the general case is given in [18].

Our predictions for $\sigma_{\hat{\mu}}^2$ can be used to characterize and validate simulations such as [28] which reconstruct the HD curve. Indeed, the $N_{\text{freq}}$ scaling behavior we describe has been observed in simulations: see [29], Eq. (24), and [30], Fig. 9.

Bayesian reconstruction of the HD correlation, starting from PTA data, produces posterior probability distributions for $\mu(\gamma)$ and $c_l$. Because it makes optimal use of all information, with reasonable priors, we expect the corresponding variances to be in good agreement with our frequentist predictions, particularly if noise-marginalized methods [25] are employed.

*Acknowledgments*—We thank Arian von Blankenburg, Serena Valtolina, and the anonymous referees for helpful corrections, comments, and suggestions. J. D. R acknowledges financial support from NSF Physics Frontier Center Award PFC-2020265 and start-up funds from the University of Texas Rio Grande Valley.